\documentclass{DISproc}

\begin{document}
%------------------------------------
\title{GPDs at an EIC}

%for single authors the superscripts are optional
\author{{\slshape Salvatore Fazio$^1$ %%, Ruth Miller$^2$
}\\[1ex]
$^1$Brookhaven National Laboratory, PO Box 5000, 11973 Upton NY, USA\\
%%$^2$CERN, 1211 Gen\`eve 23, Switzerland 
}

% please enter the contribution ID for the DOI
\contribID{xy}

\doi  % if there is an online version we will register DOIs

\maketitle

\begin{abstract}
The feasibility for a measurement of the exclusive production of a real photon, a process although known as
Deeply Virtual Compton Scattering (DVCS) at an Electron Ion Collider (EIC) has been explored. 
%% Such a new facility will
%% require the design and construction of a new optimized detector profiting from the experience gained from
%% electron-proton colliders like at the experiments H1 and ZEUS at DESY-HERA. In particular, eRHIC is a
%% machine designed to collide an electron beam with energies ranging from 5 GeV up to 30 GeV with the RHIC
%% hadron beams (protons (100 -250 GeV) and nuclei (á 100 GeV)) at varying center-of-mass energies. 
DVCS is universally believed to be a golden measurement toward the determination of the Generalized Parton Distribution (GPDs) functions. The high luminosity of the machine, expected in the order of $10^{34}cm^{-2}s^{-1}$ at
the highest center-of-mass energy, together with the large resolution and rapidity acceptance of a newly designed dedicated
detector, will open a opportunity for very high precision measurements of DVCS, and thus for the determination of GPDs, providing an important tool
toward a 2+1 dimensional picture of the internal structure of the proton and nuclei. 
%The huge impact such measurements would have on the determination of GPDs will be discussed.

%%  Place your abstract here. It should not exceed 100 words.  Please do
%%  not modify the style of the paper.  In particular, do not change
 %% width and height of the text and observe the page limits. Please
 %% do not use footnotes in the abstract or title.
\end{abstract}

\section{GPDs ans DVCS}

In order to open a new window into a kinematic regime that allows the systematic
study of quarks and gluons, the worldÕs
most versatile nuclear microscope, the Electron Ion Collider (EIC), has been proposed. With its wide range
in energy, nuclear beams and high luminosity, the EIC will
offer an unprecedented opportunity for discovery and precision measurements, allowing us to study 
the momentum and space-time distribution of gluons and sea quarks
in nucleons and nuclei.

One of the main goals of an EIC will be a precise determination of the Generalized Parton
Distribution functions (GPDs), which lead to a 2+1 dimensional imaging of the protons/nuclei
in the impact parameter space. GPDs are functions describing the distribution of quarks and
gluons in the nucleon with respect to both position and momentum. 
%The concept of GPDs has revolutionized the way scientists think about the structure of the nucleon, leading to
% completely new methods of Òspatial imagingÓ of the nucleon in the form of genuine 3-dimensional images. 
Moreover, GPDs allow us to study how the orbital motion of quarks in the
nucleon contributes to the nucleon spin - a question of crucial importance for nucleon
structure.

It is universally believed that the golden measurement toward the determination of the whole
set of GPDs is Deeply Virtual Compton Scattering (DVCS), which is the exclusive production of a
real photon. This process is sensitive to both quarks and gluons and, unlikely the exclusive
production on Vector Mesons, it is not affected by the uncertainty on the VM wave-function.
Furthermore it shows a very clean experimental signature consisting of two clusters in the
calorimeter with a track matching one of the clusters and a leading proton eventually measured
in the forward detectors (Roman Pot spectrometer).

The important observables sensitive to the GPDs are the differential cross section as a function
of the four-momentum transfer at the proton vertex, $|t|$, and the charge- and spin- asymmetries.
For the purpose of the cross section measurement it is important to remove from the signal the
background coming from the Bethe-Heithler (BH) events. The latter is essentially a QED process, known to an uncertainty in the order of $\sim3\%$ coming from the uncertainty on the proton form factor,
with the same final state topology of DVCS and can be subtracted from the signal by the means of MC
technique.
%Nevertheless, even if its cross section is well known, there is still an
%uncertainty of at minimum $\sim3\%$ coming from the uncertainty on the proton form factor.
Thus, especially for an EIC where systematics will dominate the measurements, it is
important to minimize the BH contribution, particularly at low energy configurations where
BH tends to dominate over the DVCS.
\begin{figure}[t]
  \centering
  \includegraphics[width=0.6\textwidth]{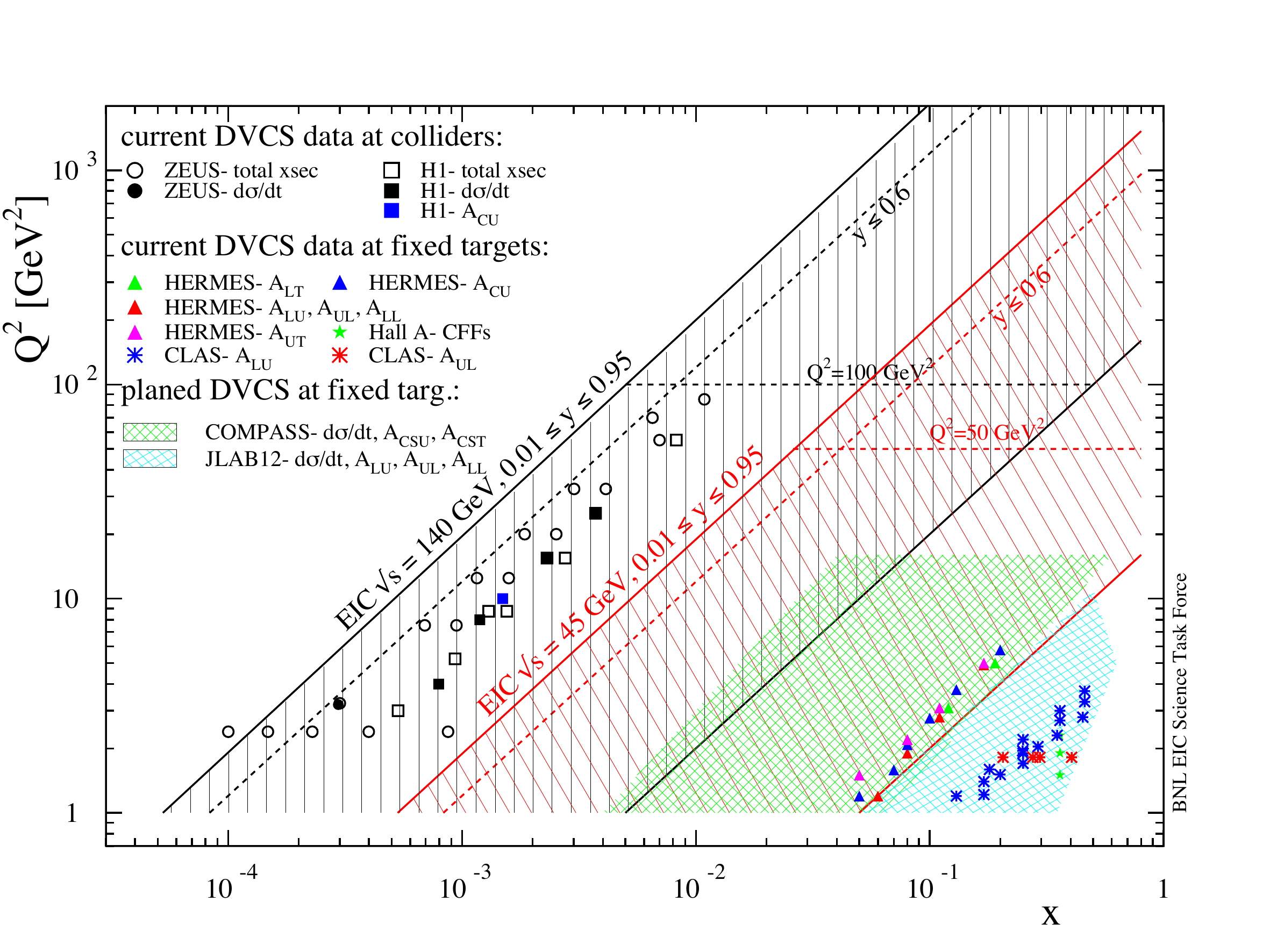}
  \caption{The EIC phase space for stage-1 (red) and stage-2 (black) compared with available data and expected future experiments.}
  \label{Fig:Phase_space}
\end{figure}
Presently available DVCS measurements provide some limited
information on GPDs and more precise data, in a wider phase space and including transversely polarized target spin asymmetry, are required to pin them down. For more informations read D. Mueller contribution in this book of proceedings.
New fixed target measurements
are planned at COMPASS II using a polarized muon beam, extending HERMES kinematics to
lower $x_{Bj}$, and at JLAB@12 GeV, see Fig.~\ref{Fig:Phase_space}. EIC will cover a much larger  phase-space and help
quantify QCD phenomena at small $x_{Bj}$~\cite{INT}. An access to GPDs requires a large
data set with small errors. In the following we would like to illustrate the potential of an EIC
machine for DVCS studies.

\section{MC simulation}

The Monte Carlo generator used for the present study is MILOU~\cite{MILOU}, which simulates both
the DVCS and the BH processes together with their interference term.
The simulated $Q^2$ and $x_{Bj}$ range corresponds to the phase space achievable with an EIC. The
electron and proton beam-energy configuration considered for the present study are: $5\text{x}100~GeV$
(stage-1) and $20\text{x}250~GeV$ (stage-2), as in~Fig.~\ref{Fig:Phase_space}.

The BH contamination has been investigated for each $Q^2$, $x_{Bj}$, $|t|$ bin as a
function of $y$. After all BH suppression criteria have been applied it was found that for stage-2 
the BH contamination grows from negligible (at low-$y$) to about $70\%$ at $y
\sim 0.6$. For stage-1, the BH contribution grows faster and can be dominant at large-$y$ depending on 
the $x_{Bj}$-bin, nevertheless most of the statistics at this low
center-of-mass energy is contained in the safe region: $y < 0.3$. It is then crucial to have a
detector which makes the experimentalists fully capable to apply all the selection criteria
required for a BH suppression (tracker: excellent angular resolution, em Cal: fine granularity and goos resolution at lower energies).

The data coming from MC simulation have been used as mock data to measure the $|t|$-differential cross section and the charge- and spin-asymmetries. Results are based on the EIC version in consideration at BNL and known as eRHIC. Simulated data samples correspond to a luminosity of $100~fb^{-1}$ for stage-2 and $10~fb^{-1}$ for stage-1 configurations, both corresponding to approximately 1 year of data taking assuming a 50\% operational efficiency. Tha variables have been smeared according to the expected resolution. A logarithmic fine binning of $x_{Bj}$ and $Q^2$ has been applied, whereas in the case of the cross section measurement, $|t|$ has been binned three times larger than the expected resolution form the Roman Pot spectrometer, which can measure proton momentum in the range $0.03<|t|<0.88~GeV^2$.  For large values of $|t|>1.0~GeV^2$ the proton can be measured in the main detector.

\begin{figure}[htbp]
  \centering
  \includegraphics[width=0.9\textwidth]{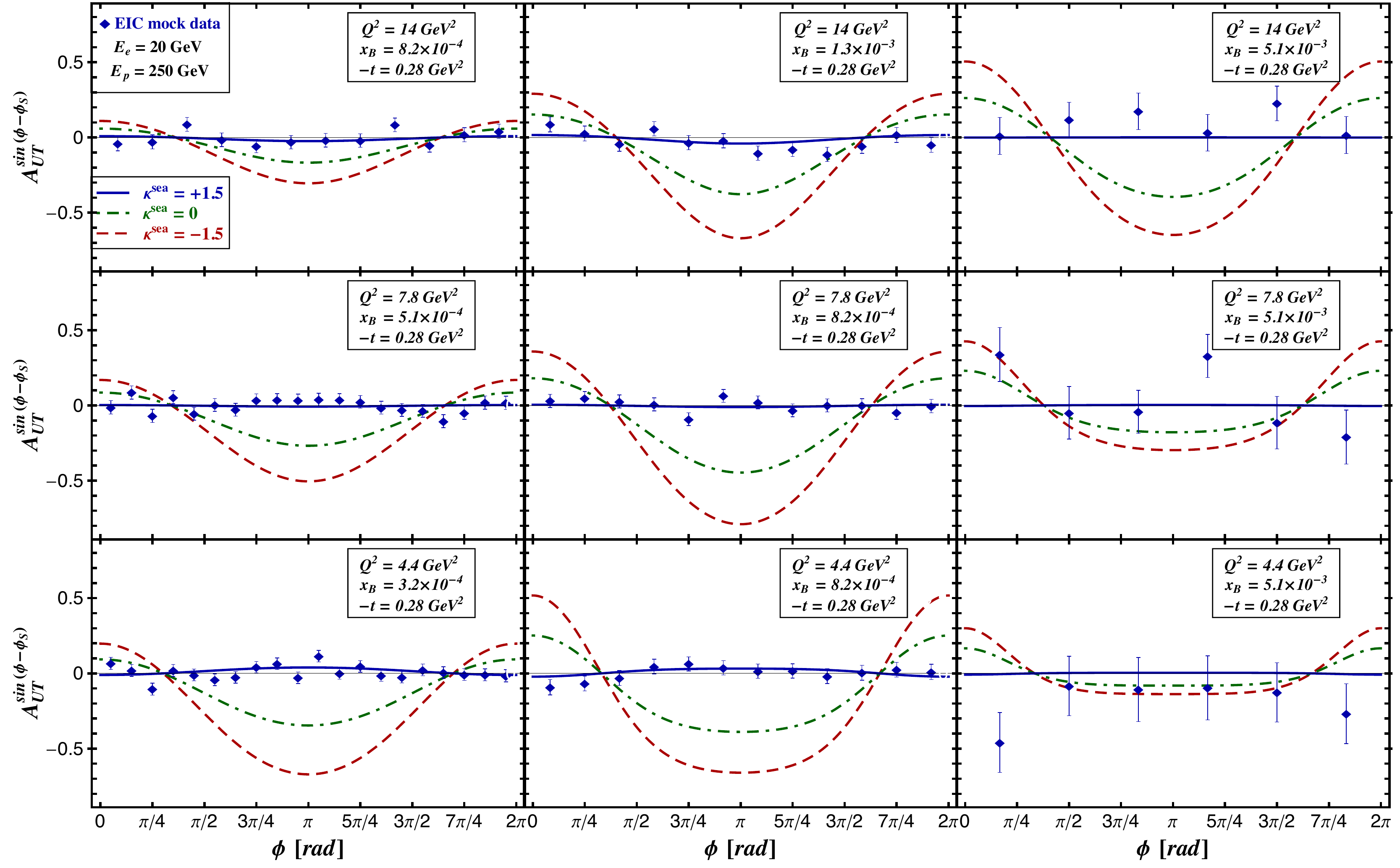}
  \caption{Transverse target-spin asymmetry uncertainties for EIC stage-2 mock data ($diamonds$) compared to theory model with large positive ($solid$), vanishing ($dot-dashed$), and large negative ($dashed$) $E^{sea}$ contributions.}
  \label{Fig:TSA}
\end{figure}  
  
The simulation proves that an EIC can perform accurate measurements of cross sections and asymmetries in a very fine binning and with a statistical acceptance often as low as a few percent. For more details and figures see the D. Mueller's and M. Diehl' contributions. 
This implies that the measurement is actually limited by systematics. For the purposes of the present study, a systematic
uncertainty of 5\% has been assumed, based on the experience achieved at HERA and the expected
acceptance and technology improvements of a EIC new detector. The overall systematic uncertainty
due to the uncertainty on the measurement of luminosity was not considered here,
since it simply affects the normalization of the cross section measurement.
As an example of the precision achievable at an EIC, Fig.~\ref{Fig:TSA}  shows the expected uncertainty for the transverse target-spin asymmetry ($A_{UT}$) as a function of the azimuthal angle $\phi$ between the production and the scattering planes for a particular $x_{Bj}, Q^2, |t|$ bin, compared to theoretical expectations.

\begin{figure}[htbp]
  \centering
  \includegraphics[width=0.65\textwidth]{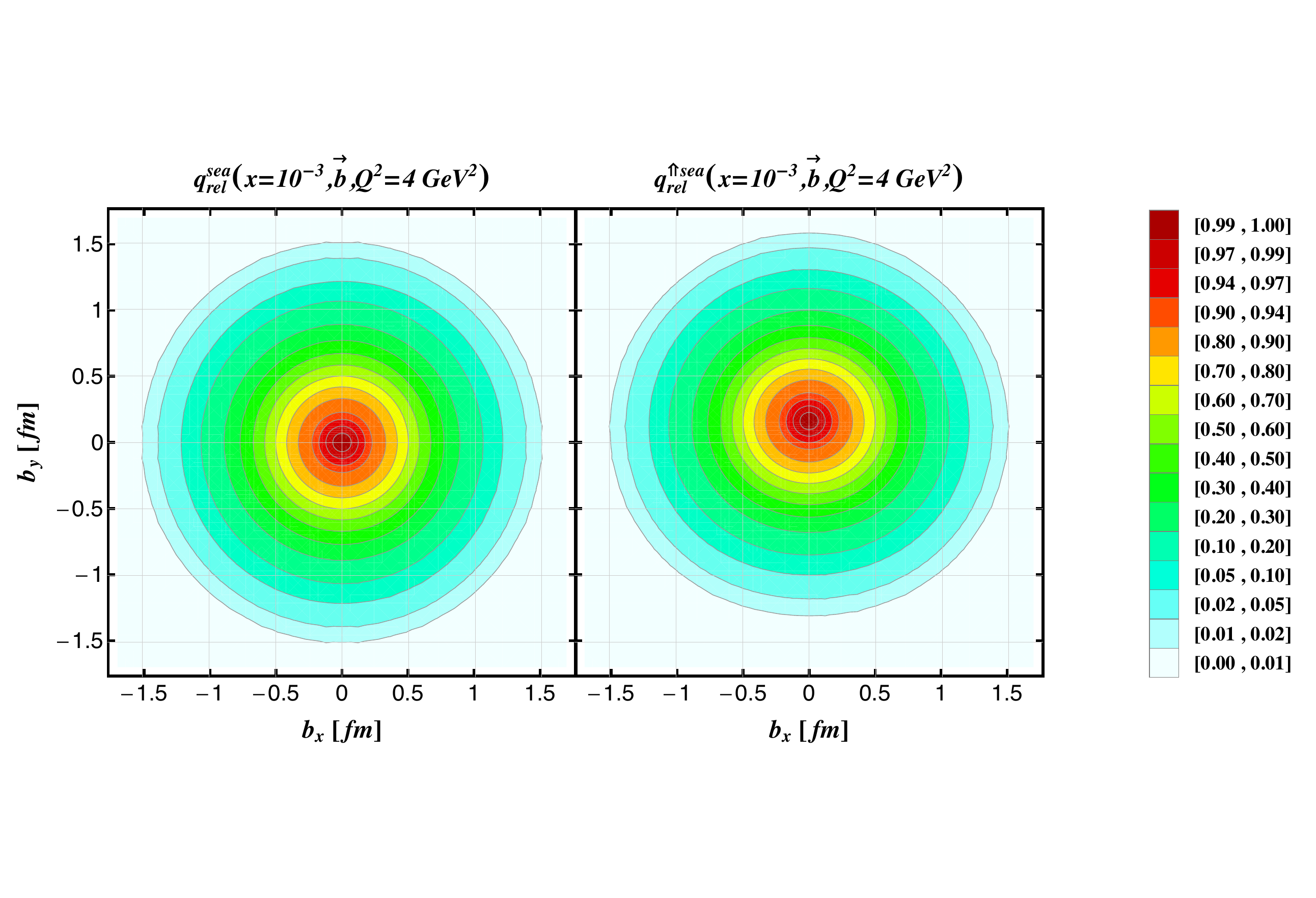}
  \caption{Tomographical picture of the sea-quuarks distribution in the impact parameter space for an unpolarized ($left$) and a polarized ($right$) proton beam.}
  \label{Fig:GPD-2d}
\end{figure}

Mock data have been then used, together with the data presently available, to constrain the GPDs. It was found that an EIC would have a great impact on the knowledge of GPDs, especially of GPD $E$, which at the moment remains unconstrained. For more details and discussion see D. Mueller's contribution. Fig.~\ref{Fig:GPD-2d}  shows an example of a tomographic picture of the see-quarks distribution in the nucleon in the impact parameter space, as resulting from EIC mock data analysis, for a particular  $x_{Bj}, Q^2$ bin, for the case of an unpolarized and a polarized  target-beam.

\section{Conclusions}

To conclude, an EIC will be a unique facility to study DVCS with high
precision and accuracy. The very high luminosity of the machine together
with the precision of $|t|$ measurement from a dedicated spectrometer and
the tracker acceptance at large rapidities opens the possibility of a fine binning in $Q^2$ and $x_{Bj}$ and $|t|$ with a very low uncertainty. 
This will give a precious contribution to the
GPDs extraction and will help to discriminate among different theoretical
models.

%\section{Bullshits}
%By default the following extra packages are included:
%\begin{description}
 % \setlength{\parskip}{0pt}\setlength{\itemsep}{0pt}
%\item[graphicx:] tor including figures;
%\item[wrapfig:] to wrap text round figures and tables;
%\item[rotating:] to rotate elements of the text, e.g.\ tables;
%\item[amssymb, amsmath, array:] extra math environment options and macros;
%\item[hepunits:] package to help formatting units;
%\item[csquotes:] package to help with consistent and correct usage of quotes;
%\item[booktabs:] improve the look of tables
%\item[hyperref:] enable clicking on cross-references and URLs in the text.
%\end{description}

%If some packages already included in the template are not installed on 
%your machine contact the system administrator or comment out the relevant
%package, e.g.\ \texttt{wrapfig} or \texttt{rotating}.

%If you have an old version of \LaTeX\ or want to use a system such as
%\texttt{lxplus} at CERN that does not have all the packages, you can
%download \texttt{hepunits.tar.gz}, \texttt{csquotes.tar.gz} and \texttt{easytoolbox.tar.gz}
%from the DIS 2012 web page. These files should be unpacked into your
%\texttt{\textasciitilde /texmf/tex/latex} directory. 

%If you install \texttt{csquotes} and \texttt{etoolbox} and PDF\LaTeX\
%still complains (e.g.\ on DESY DL5 machines) try \texttt{pdfelatex} instead. 

%\section{Bibliography}
 
% ****************************************************************************
% BIBLIOGRAPHY AREA
% ****************************************************************************

{\raggedright
\begin{footnotesize}
% IF YOU DO NOT USE BIBTEX, USE THE FOLLOWING SAMPLE SCHEME FOR THE REFERENCES
% ----------------------------------------------------------------------------

% ----------------------------------------------------------------------------

% IF YOU USE BIBTEX,
% - DELETE THE TEXT BETWEEN THE TWO ABOVE DASHED LINES
% - UNCOMMENT THE NEXT TWO LINES AND REPLACE 'smith_joe.bib' WITH YOUR
%   FILE(S)

% \bibliographystyle{DISproc}
% \bibliography{smith_joe.bib}

\begin{thebibliography}{99}
%------- replace following references ;-)
\bibitem{INT} D. Boer et al.,  ``Gluons and the quark sea at high energies: Distributions, polarization, tomography'', 2011,
{{\ttfamily arXiv:1108.1713}}.
\bibitem{MILOU} E. Perez, L. Schoeffel, and L. Favart, (2004), {{\ttfamily hep-ph/0411389}}.
%\bibitem{parton_qed} A.~D.~Martin {\it et~al.} Eur. Phys. J. {\bf C39} (2005) 155.
%\bibitem{DVCS1}S.~Friot, B.~Pire and L.~Szymanowski. Phys. Lett. {\bf B645} (2007) 153.
%\bibitem{DVCS2}D.~Hasell, R.~Milner and K.~Takase. AIP Conf. Proc. {\bf 588} (2001) 187.
%\bibitem{DVCS3}M.~Krawczyk and A.~Zembrzuski. Phys. Rev. {\bf D57} (1998) 10.
%\bibitem{pomeron1}R.~Brower and C.~Tan. PoS LAT2005 (2006) 279.
%\bibitem{pomeron2}J.~P.~Guillaud and A.~Sobol,
%  ``Perspectives of the study of double Pomeron exchange at the LHC'', in
%  {\it 11th Lomonosov Conference on Elementary Particle Physics}, Moscow, Russia, 2003.
%\bibitem{Gogitidze:2007du} {\bfseries H1} Collaboration, N.~Gogitidze, ``Prompt photons and particle momentum distributions at hera''. hep-ex/0701033, 2007.
%\bibitem{H1}H1 Collab., N.~Gogitidze {\it et~al.}, ``{Prompt photons and particle momentum distributions at HERA}'', 2007. 
%\href{http://arxiv.org/abs/hep-ex/0701033}{{\ttfamily hep-ex/0701033}}
\end{thebibliography}
\end{footnotesize}
}

% ****************************************************************************
% END OF BIBLIOGRAPHY AREA
% ****************************************************************************

\end{document}